\documentclass[prb,twocolumn,showpacs,a4]{revtex4}
\usepackage{graphicx}
\usepackage{epsfig}
\usepackage{amssymb,latexsym,amsmath}

\usepackage{ulem}
\usepackage{color}

\begin{document}
\title{ \Large{ \bf Conditions for requiring nonlinear thermoelectric transport theory in nanodevices } }

\author{J. Azema, P. Lombardo, and A.-M. Dar\'e}\email{Anne-Marie.Dare@univ-amu.fr}
\affiliation {Aix-Marseille Universit\'e, CNRS, IM2NP UMR 7334, \small \it  13397, Marseille, France}

\date{\today}

\begin{abstract}

In this paper, we examine the conditions under which the nonlinear transport theory is 
inescapable, when a correlated quantum dot is symmetrically coupled to two leads submitted to temperature and voltage biases.
By detailed numerical comparisons between nonlinear and linear currents,  
we show that the claimed nonlinear behavior in a temperature gradient for the electric current is not so genuine, and 
the linear theory made at the operating temperature $\overline{T}= (T_H+T_C)/2$ is unexpectedly robust. This is demonstrated 
for the single impurity  Anderson
model, in different regimes: resonant tunneling, Coulomb blockade, and Kondo regimes.

\end{abstract}
\maketitle

\section{Introduction}

In the wide and strategic search for energy harvesting, nanoscale
thermoelectric engines have
become promising devices to improve the energy conversion efficiency.
In this context many experimental realizations, from carbon nanotubes \cite{Small03} to molecular
junctions \cite{Reddy07} and quantum dots \cite{Staring93,Scheibner05,Scheibner07,Svensson12,Svensson13,Thierschmann13}, have been proposed and characterized.
Meanwhile, the theoretical works have not been overlooked.  This
field is very active, as evidenced by the numerous review articles
\cite{Dubi11,Zebarjadi12,Sothmann14,Wang08,Zimbov11,Giazotto06,Goupil11}.

Since a few years --it presumably began with a famous publication by Reddy {\it et al.} \cite{Reddy07}-- numerous studies highlight the nonlinear properties of nanodevices subjected
 to some temperature gradient \cite{Dubi11,Kirchner12,LeHur13,Grifoni12,Leijnse2010,Hershfield13,Withney13,Withney13ter,Sanchez2013,Lopez2013,Sierra2014}.
The nonlinearity of devices submitted to voltages has also been raised, probably beginning with Christen and B\"uttiker\cite{Christen96}.
It seems to us that 
the debate between linear and nonlinear behaviors in thermoelectric properties of nanoscopic devices 
covers two different aspects.
The first one is practical: considering a device submitted to some temperature and/or voltage gradient,
it is natural 
to ask whether the measured property, electric current, for example, is an affine function 
of these experimentally applied biases. 
The answer to this question is expected to depend on the particular way to apply the biases,
this is illustrated in Sec. VI B.

The other aspect relies on the following question:  to account for some observed property at finite biases, 
is it necessary to pull out the heavy artillery of out-of-equilibrium physics (quantum theories of transport ideally 
including Coulomb correlations), or can the
evaluation of the transport coefficients be sufficient to understand and to give a quantitative and reliable estimation of the property?
When this second scenario applies, the linear transport theory which depends only on equilibrium properties 
(much less involving than nonequilibrium calculations) can be used.

Sometimes these two different aspects are ambiguously mixed up. We try to disentangle these two facets 
in the particular case of a correlated, spin-degenerate two-orbital quantum dot, coupled to noninteracting 
leads.
To do so, we quantitatively compare the results of a general {\it a priori} nonlinear electric current calculation, in presence of thermal and voltage 
biases, to an evaluation of the current based on transport coefficients. 
We thus delimit the region where the out-of-equilibrium calculations are inescapable. 
We also show that linear transport
theory and linear behavior are not synonymous. 
Indeed, when out-of-equilibrium physics is necessary, under certain circumstances, an affine relation between current and thermal bias may be obtained,
while conversely, a non monotonous relation may be observed, which can nevertheless be quantitatively reproduced within the linear transport theory.

Only few approaches enable to make out-of-equilibrium calculations including properly Coulomb interactions. Undeniably, to handle at the same time out-of-equilibrium physics and correlation physics is a challenge. 
This difficulty is a strong argument for trying to take advantage at best of the linear transport approach.
We use such a theory, namely the generalized Keldysh-based out-of-equilibrium non crossing approximation (NCA)~\cite{WM94,HKH98},
which has been shown to reliably describe transport properties down to temperatures of the order of a fraction of the Kondo temperature~\cite{Aligia2012}.
Concerning thermoelectric properties, NCA provides a description~\cite{Azema12} of transport through a correlated quantum dot in the Kondo regime, which is consistent with numerical renormalization group results~\cite{Costi10}.

Within the Meir-Wingreen-Landauer formalism \cite{WM94}, the electric current flowing from left  to right, can be expressed as
\begin{equation}
I = \frac{2e}{h}\int \left[f_{\rm L}(\varepsilon)-f_{\rm R}(\varepsilon)\right]\tau(\varepsilon)\,{\rm d}\varepsilon \ ,
\label{Landauer}
\end{equation}
in terms of a transfer function $\tau(\varepsilon)$,
times the lead Fermi function difference; with
$f_{\alpha}(\varepsilon)\equiv  \Bigl( e^{(\varepsilon-\mu_{\alpha})/(k T_{\alpha})}+1\Bigr)^{-1}$, 
$h$ and $k$ denote respectively the Planck and Boltzmann constants, and $-e$ the electronic charge. 
The formula (\ref{Landauer}) is exact for electric transport through some central region connected to uncorrelated reservoirs, in case of proportional left and right couplings. [See the next section for a definition of $\tau(\varepsilon)$ for the model under scrutiny.]

In case of a rigid transfer function with respect to temperature and voltage,
 the debate between linear and nonlinear transport approaches is readily settled: just by examining the series expansion of the Fermi function difference, which 
weights the transfer function. 
It is clear that the best point around which making the expansion, is the average temperature and chemical potential, $\overline{T}$ and $\overline{\mu}$. 
If the difference between $(f_L(\varepsilon) - f_R(\varepsilon))$ and its first-order series expansion is slight for the energy range corresponding to nonvanishing rigid $\tau(\varepsilon)$, the linear response approach will be satisfactory.
From the Fermi function difference, the criteria for validity of the linearization are simply set by the ratios ${V_b} / {\overline{T}}$ and ${\Delta T} /{\overline{T}}$, where $V_b$ and $\Delta T$ are the electrostatic and temperature biases. 
In this paper, we show that, surprisingly, despite a nonrigid transfer function, as obtained in the Anderson model, and for a wide range of parameter values, the criteria for validity of the linear theory are essentially the same! 
We will also show that this result does not 
strongly
rely on the wide-band-limit assumption. 

The paper is organized as follows. After an introduction to the model, and a quick glance at the Fermi function difference and its series expansions, we present our findings in two different parameter areas: first for a quasiresonant tunneling case, second in a regime where the Kondo physics enables to circumvent the Coulomb blockade and restores the transport through the dot. More precisely, in both cases, we present the out-of-equilibrium transfer function and nonlinear electric current that are compared to their equilibrium and linear counterparts. We thus outline the bias region where the nonlinear calculations are essential. 
We also calculate in the first case the linear and nonlinear thermopower.
We end the paper with some miscellaneous points, a calculation of nonlinear current beyond the wide band limit,  as well as an example that shows that linear theory and affine relation should not be confused.

\section{Model}

The device we consider consists of a central dot coupled to two uncorrelated leads.  Retaining two relevant orbitals in the dot, the system corresponds to an Anderson model, with a Hamiltonian given in standard notation by \cite{A61}
\begin{eqnarray}
\label{Hamiltonian}
H&=&
\epsilon_{0} \sum_{m,\sigma} c_{m \sigma }^{\dagger } c_{m \sigma } 
\,+\,\frac{U}{2}\sum_{{(m,\sigma)\neq}\atop {(m^\prime,\sigma^\prime})} n_{m\sigma} n_{m^\prime \sigma^\prime }
\nonumber  \\
&+& \sum\limits_{{\alpha\in\{L,R\}}\atop {k,m,\sigma}}\epsilon_{\alpha k}\,a_{\alpha k m\sigma }^{\dagger} a_{\alpha k m\sigma}
\nonumber \\
&+& \sum\limits_{{\alpha\in\{L,R\}}\atop {k,m,\sigma}} 
\left( t_{\alpha} c_{m\sigma }^{\dagger} a_{\alpha k m\sigma}+t_{\alpha}^*a_{\alpha k m\sigma }^{\dagger } c_{m \sigma }\right)
\,\mbox{.}
\end{eqnarray}
The first line concerns the  doubly degenerate ($m=1,2$) orbitals with on-site Coulomb repulsion $U$, the second line describes the left (L) and right (R) leads, and the last line accounts for the tunneling between dot and leads which conserves orbital and spin quantum numbers.

The transfer function appearing in Eq.(\ref{Landauer}) is defined by
$\tau(\varepsilon)=\pi A(\varepsilon)\Gamma(\varepsilon)/4$, where $\Gamma(\varepsilon)$ is  determined by the hybridization strength and the lead densities of states:
$\Gamma(\varepsilon)= \Gamma_L(\varepsilon)+\Gamma_R(\varepsilon)$, with $\Gamma_{\alpha}(\varepsilon)= \pi t_{\alpha}^2 N_{\alpha}(\varepsilon)$. $N_{\alpha}(\varepsilon)$ is the $\alpha$-lead spin-summed density of states. 
Unless explicitly stated we restrict our calculations to the wide band limit, as a consequence $\Gamma_{\alpha}(\varepsilon)$ is nearly independent of energy  (we use a Gaussian function for $\Gamma_{\alpha}(\varepsilon)$: defining $\Gamma= \Gamma(\overline{\mu})$, henceforth the energy unit, the full width at half maximum of the Gaussian is $169 \Gamma$).
Finally, defined in terms of the retarded Green's function, $A(\varepsilon)=-\frac{1}{\pi}{\rm Im}[\sum_{m,\sigma}G^{r}_{m\sigma}(\varepsilon)]$ is the dot spectral density, summed over orbital and spin degrees of freedom. This spectral density is evaluated within a generalized Keldysh-based out-of-equilibrium NCA~\cite{WM94,HKH98}.

We only consider a symmetrically coupled quantum dot: $\Gamma_L = \Gamma _R$. 
The electric bias $V_b$ will be applied symmetrically around a fixed $\overline{\mu}$ (henceforth our energy origin): 
$\mu_{L(R)} = \mp e V_b/2$, therefore the dot 
level is
bias-independent. If the capacitive couplings between dot and both leads are equal ($C_L = C_R$), 
and if the capacitive coupling between dot and gate can be neglected ($ C_G \ll C_L +  C_R $), the symmetric way to apply electrostatic voltages is equivalent to applying them dissymmetrically, for example, with one lead grounded, 
and a voltage drop 
of $V_b/2$ affecting the dot, as routine in experimental setups.

In the out-of-equilibrium case, the spectral function $A(\varepsilon)$ does not only depend on the Hamiltonian parameters, but also on the lead temperatures and voltages. 
An extreme example is 
the Kondo resonance, which is known to be very sensitive to voltages and may disappear at low bias as shown in Ref. \cite{WM94}.
For our choice of energy origin, 
the voltages enter $A(\varepsilon)$ only through their difference $V_b=V_L - V_R$. There is no such simplification for temperatures. $A(\varepsilon)$ is {\it a priori}  a function of both independent variables $T_L$ and $T_R$, or equivalently a function of the average temperature $\overline{T} =\frac{T_L+T_R}{2}$, and of the difference $\Delta T = T_L -T_R$.
 
 \section{Series expansion of the Fermi function difference}
 
\begin{figure}[h!]
            \includegraphics[width=0.5\textwidth]{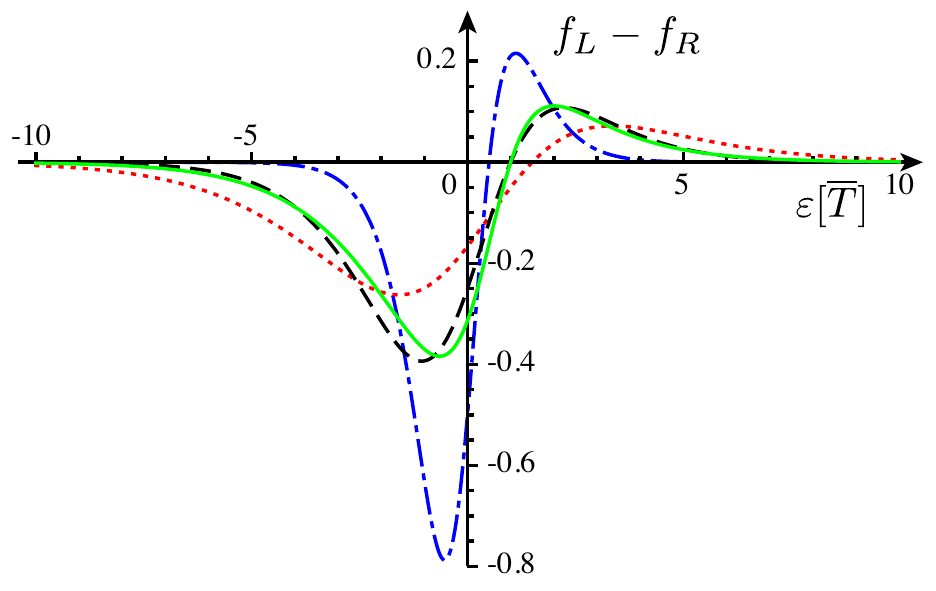}
    \caption{Fermi function difference $(f_L - f_R)$ as a function of energy, and its various first-order series expansion for $V_b = \Delta T = T_L -T_R = \overline{T}$. The energy unit is $\overline{T}$.
Exact difference:  green line; series expansion at $\overline{T}$: black dashed line; series expansion at $T_R=\overline{T}-\Delta T/2$: blue dot-dashed line; series expansion at $T_L= \overline{T}+\Delta T/2$: red dotted line.}
    \label{DL}
\end{figure}

 As an introduction, let us rapidly examine the series expansion of the Fermi function difference that weights $\tau(\varepsilon)$ in the Meir-Wingreen expression (\ref{Landauer}).
It may be instructive to have a look at Fig.~\ref{DL}, where 
the exact difference $(f_L - f_R)$ 
and its various first-order expansions are superimposed for the following parameters: $\Delta T = \overline{T}$, $ V_b = \overline{T}$ (henceforth $e=1$). 
The voltage expansion is made around $\overline{\mu}=0$, whereas we keep some freedom in the temperature one, presenting the development around the average, {\it cold }, and {\it hot} 
temperatures.
By far, the best expansion, as expected, is around $\overline{T}$. Indeed, as seen on the figure,
the Fermi function difference may be respectively significantly amplified  or eroded locally by a development around the {\it cold} or {\it hot} temperatures. 
The validity of the linearization of the Fermi function difference relies only on the ratio values ${V_b} / {\overline{T}}$ and ${\Delta T} / {\overline{T}}$. 
The agreement between fully nonlinear and linear approaches 
for quantities as the current,
may also depend significantly on the locations and widths of spectral density structures, as well as on their temperature and bias dependencies, this will be addressed in detail in the next two sections. Nevertheless, just on Fermi function expansion considerations, we cannot expect a satisfactory linearization 
for ${|V_b|} , {|\Delta T|} > {\overline{T}}$.

 \section{Quasiresonant tunneling}
 
\subsection{Out-of-equilibrium spectral function}

\begin{figure}[h!]
           \includegraphics[width=0.5\textwidth]{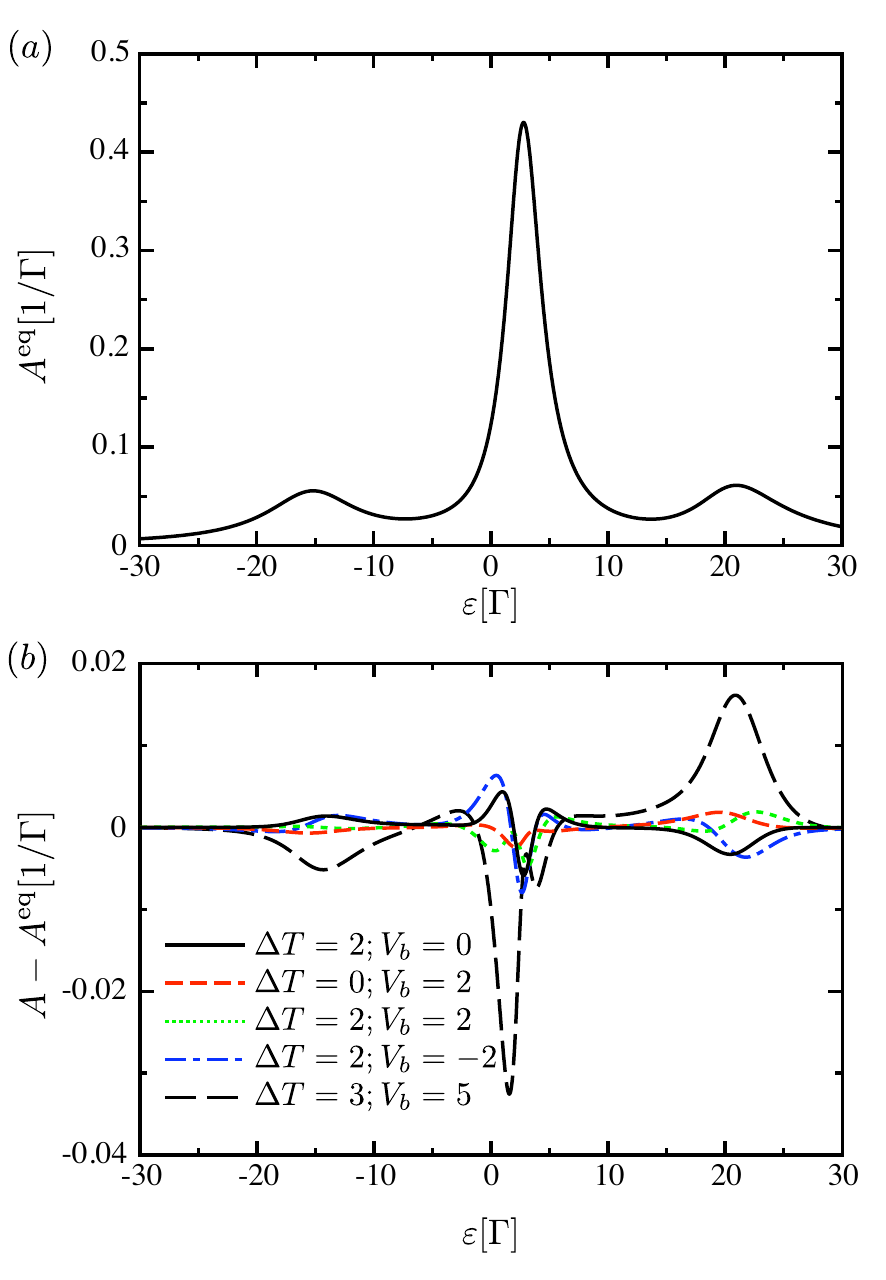}
    \caption{Spin and orbital summed spectral density function as a function of $\varepsilon$, for $\varepsilon_0 =-13.54 \Gamma$, $U =16 \Gamma$, $\overline{T}=2 \Gamma$ at equilibrium (a), and (b)  difference between nonequilibrium $A$ and equilibrium $A^{\rm eq}$  spectral densities for different thermal and electric biases in $\Gamma$ units. }
    \label{resTbar2}
\end{figure}

\begin{figure}[h!]
            \includegraphics[width=0.5\textwidth]{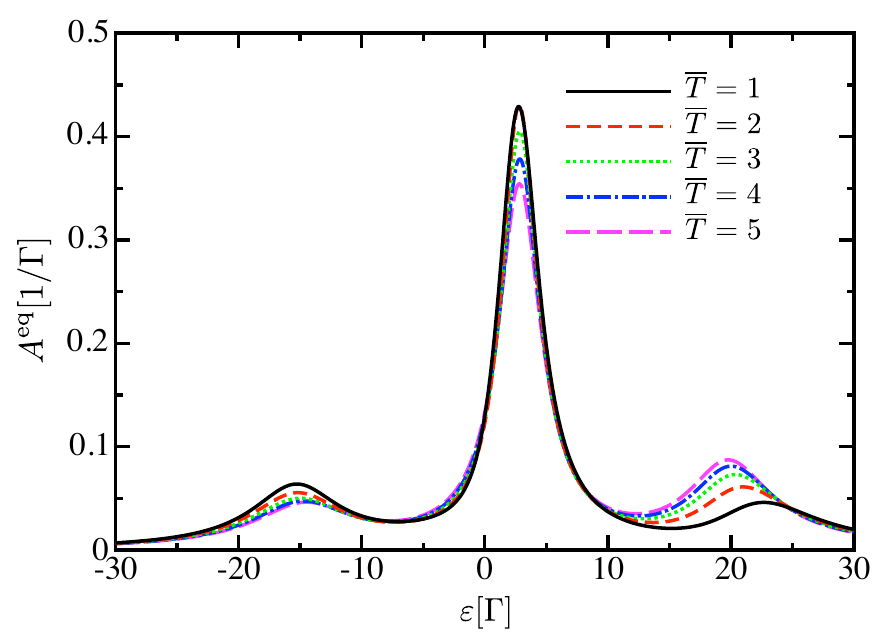}
    \caption{Equilibrium spectral function for the same parameters as in Fig.~\ref{resTbar2} and for different temperatures given in $\Gamma$ units. }
    \label{resEquil}
\end{figure}

In the out-of-equilibrium case, the transfer function acquires voltage and temperature dependencies through its relation to the spectral density function, which may be written explicitly as $A(\varepsilon, \overline{T}, \Delta T, V_b)$. By symmetry reasons it has the property $A(\varepsilon, \overline{T}, -\Delta T,  -V_b)=A(\varepsilon, \overline{T}, \Delta T, V_b)$.
In the quasiresonant regime (a regime where there is significant spectral density close to the mean chemical potential), which can be reached for example for $\varepsilon_0 =-13.54 \Gamma$ and $U =16 \Gamma$, we examine this function  in various $V_b$ and temperature circumstances.
First we choose a moderate average temperature: $\overline{T}=2\Gamma$. In Fig.~\ref{resTbar2}(a), the spectral density at equilibrium,
$A^{\rm eq} = A(\varepsilon, \overline{T}, \Delta T=0, V_b=0)$,
is displayed, and in
Fig.~\ref{resTbar2}(b), the difference between nonequilibrium and equilibrium spectral densities is shown for various thermal and electric biases. 
The difference is quite small for moderate biases, up to $\Delta T \simeq |V_b| \simeq \overline{T}$, but raises and may become significant for higher voltage or temperature differences.
However, it would be wrong to think that
this is due in part to a faint temperature dependence of $A$, as revealed in Fig.~\ref{resEquil} where the equilibrium case is plotted for various $\overline{T}$: $A$ is not at all a rigid function of temperature, and $\overline{T}$ affects not only the amplitude but also the location of the spectral density structures. 
To have a quantitative estimation, one can note that between the nonequilibrium spectral function at $(\overline{T}=2\Gamma, \Delta T=2\Gamma,V_b=0)$ and the equilibrium one at $\overline{T}=2\Gamma$, the maximum difference is about $5 \times 10^{-3} \Gamma^{-1}$, whereas the difference between the equilibrium spectral functions corresponding to the minimal and maximal temperatures, respectively $\overline{T}=\Gamma$ and $\overline{T}=3\Gamma$, can be about ten times larger.
For the shown parameters, the equilibrium spectral density is a very good approximation of the out-of-equilibrium one, as long as $|\Delta T|, |V_b| \lesssim \overline{T}$.

In the energy window displayed in previous figures, the two-orbital-degenerate spectral function exhibits three transition peaks, corresponding to on-dot charge fluctuations between respectively empty and singly occupied (peak located close to $\varepsilon_0$), singly and doubly occupied $(\varepsilon_0+U)$, and doubly and triply occupied states $(\varepsilon_0+2 U)$. The fourth peak, corresponding to the transition between three and four electrons on the dot is quite small, and is located at higher energy ($U$ farther from the third peak), such that besides its low amplitude, it will be outside the window defined by the Fermi function difference, thus it will not significantly contribute to the transport for the parameters we consider. 

\begin{figure}[h!]
         \includegraphics[width=0.5\textwidth]{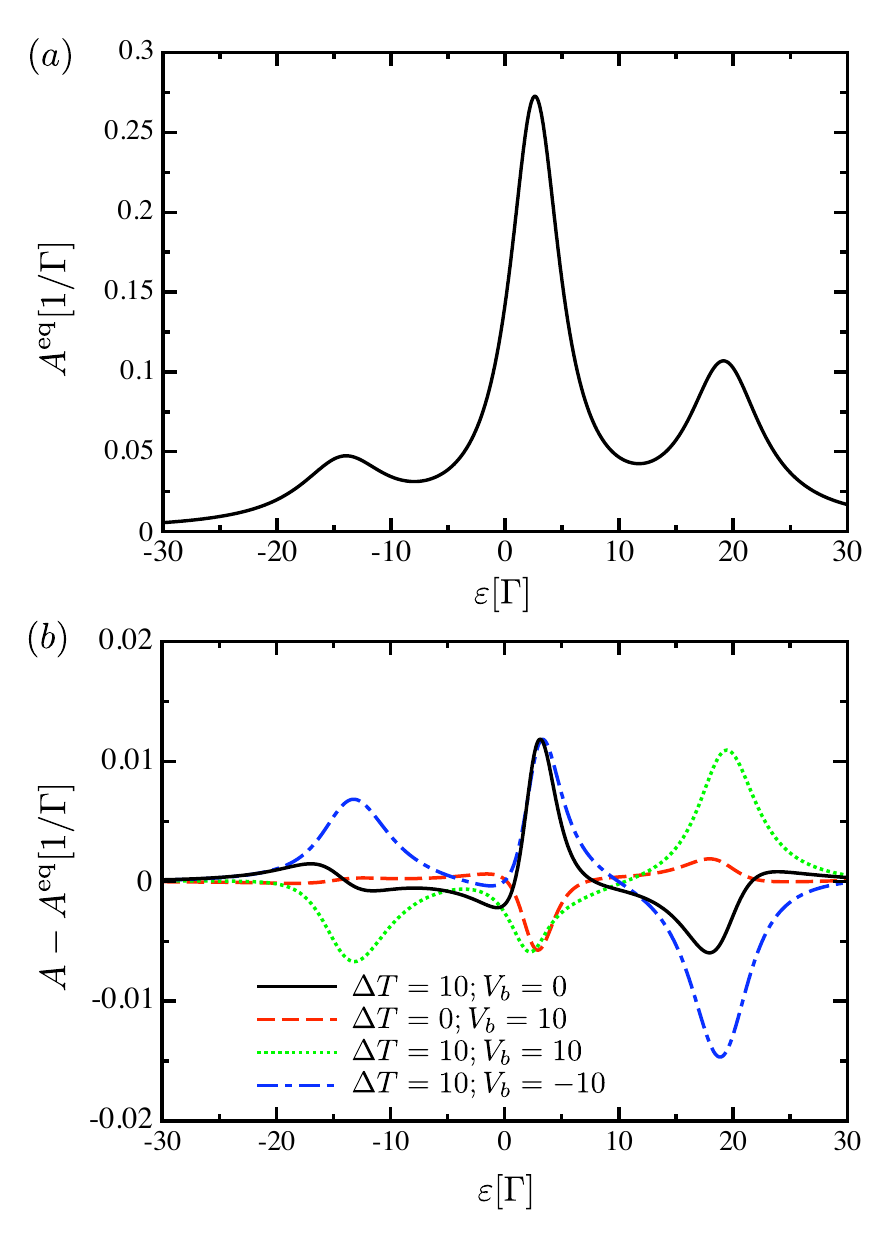}    
         \caption{Spin and orbital summed spectral density function for $\varepsilon_0 =-13.54 \Gamma$, $U =16 \Gamma$, $\overline{T}=10 \Gamma$ at equilibrium (a), and (b) difference between nonequilibrium $A$ and equilibrium $A^{\rm eq}$  spectral densities for different thermal and electric biases in $\Gamma$ units. }
    \label{resTbar10}
\end{figure}

The only ratios involved in the validity criterion of Fermi function difference expansion are $V_b / \overline{T}$ and $\Delta T / \overline{T}$. However, some other energy scale may interfere to validate or invalidate the approximation of the spectral density by its equilibrium counterpart; indeed, $\Gamma$ may also be at stake. 
This was suggested in Ref. \cite{Dubi11}
to rule out the linear response theory for $|\Delta T |> \Gamma$. We now show, and later in the Kondo regime, that $\Gamma$ does not have such an important role: the same conclusions are found for a mean temperature much higher than $\Gamma$, and will be found again in the opposite regime. The equilibrium spectral function in case of a high average temperature is shown in Fig.~\ref{resTbar10}, as well as the difference between out-of-equilibrium and equilibrium ones, for different biases. 
If the discrepancies between equilibrium and out-of-equilibrium spectral density are slightly larger than they were for a moderate $\overline{T}$, they are still modest, as long as 
$|\Delta T|, |V_b| \lesssim \overline{T}$.
A detailed examination of this figure reveals that a temperature bias is somewhat more efficient to disturb the equilibrium properties than a pure voltage one. 
Let us now turn to the current calculation itself.

\subsection{Quantitative comparisons between nonlinear and linearized currents }

In the linear transport theory, the electric current (defined  here to be positive for electrons flowing from left to right), for $V_b=V_L - V_R$ and $\Delta T =T_L -T_R$, is given by $I_L= -G V_b - S G \Delta T$. The transport coefficients, respectively, the conductance and Seebeck coefficient, are given by $G(T) = e^2 I_0(T)$ and $S(T) = - 1 / (e T)  I_1(T) / I_0(T)$ where, according to Ref. \cite{Kim03}, 
\begin{equation}
I_n(T)= \frac{2}{h}\int  \varepsilon^n   \Bigl( -\frac{\partial f}{\partial \varepsilon} \Bigr) \tau^{\rm eq}(\varepsilon)     {\rm d}\varepsilon        \ ,
\end{equation}
with $\tau^{\rm eq}(\varepsilon)$ the equilibrium transfer function.
As previously argued, to optimize the regime where the linear theory is presumed to be valid, just on Fermi function difference consideration, we shall use
 the transport coefficients evaluated at the average temperature $\overline{T}$, to calculate the linearized current.

\begin{figure}[h!]
         \includegraphics[width=0.5\textwidth]{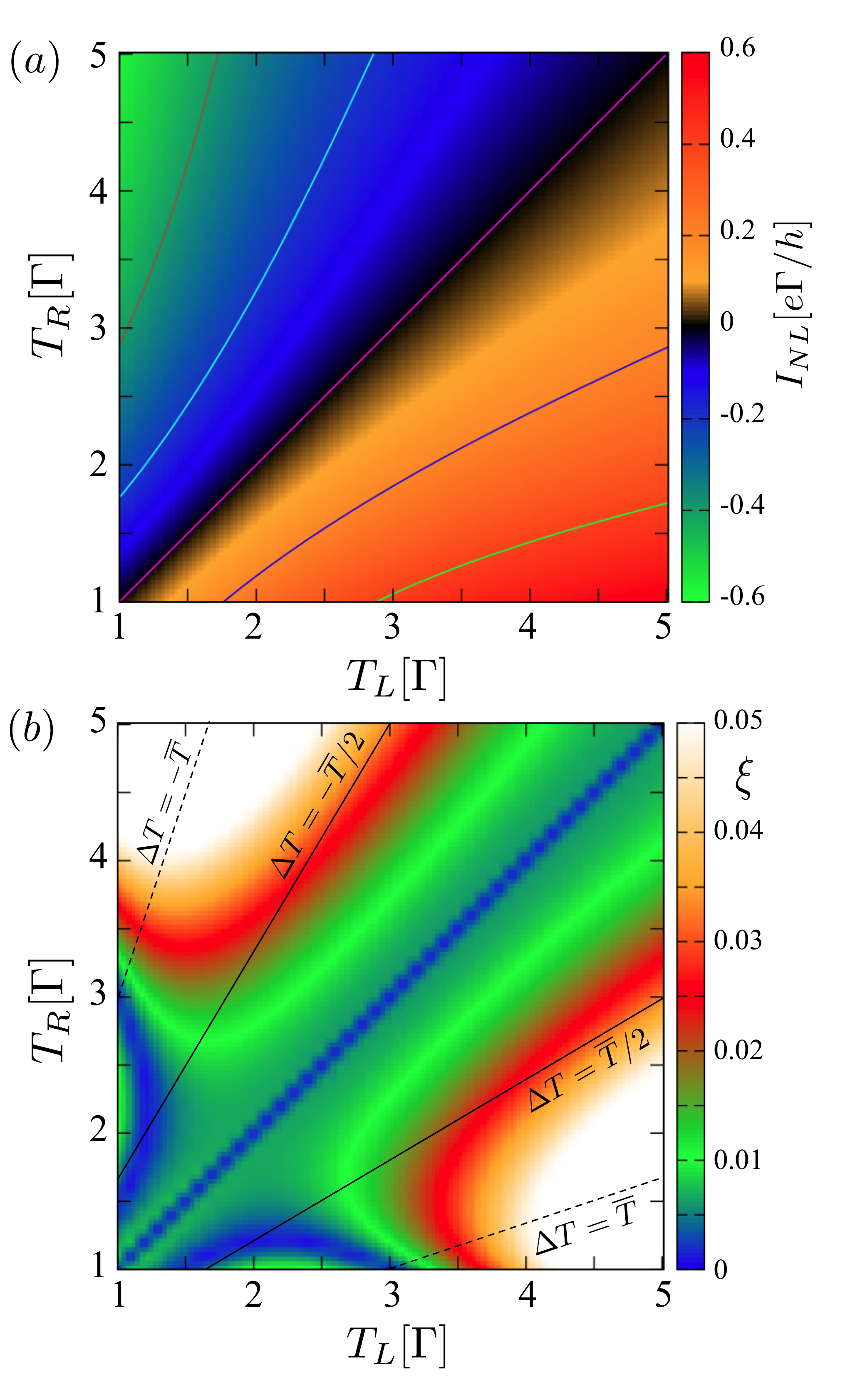}  
    \caption{Maps of nonlinear current $I_{NL}$ (a), and relative deviation to the linear expansion  $\xi=|(I_{NL}-I_L) /  I_{NL}|$ (b), for the following parameters:  $\varepsilon_0 = -13.54 \Gamma$, $U =16 \Gamma$, and $V_b=0$.
The iso-current curves in (a) are $\sim 0.2 e \Gamma / h$ apart.  See text for the lines superimposed on the graph (b).}
    \label{Vbnul}
\end{figure}

\begin{figure}[h!]
      \includegraphics[width=0.5\textwidth]{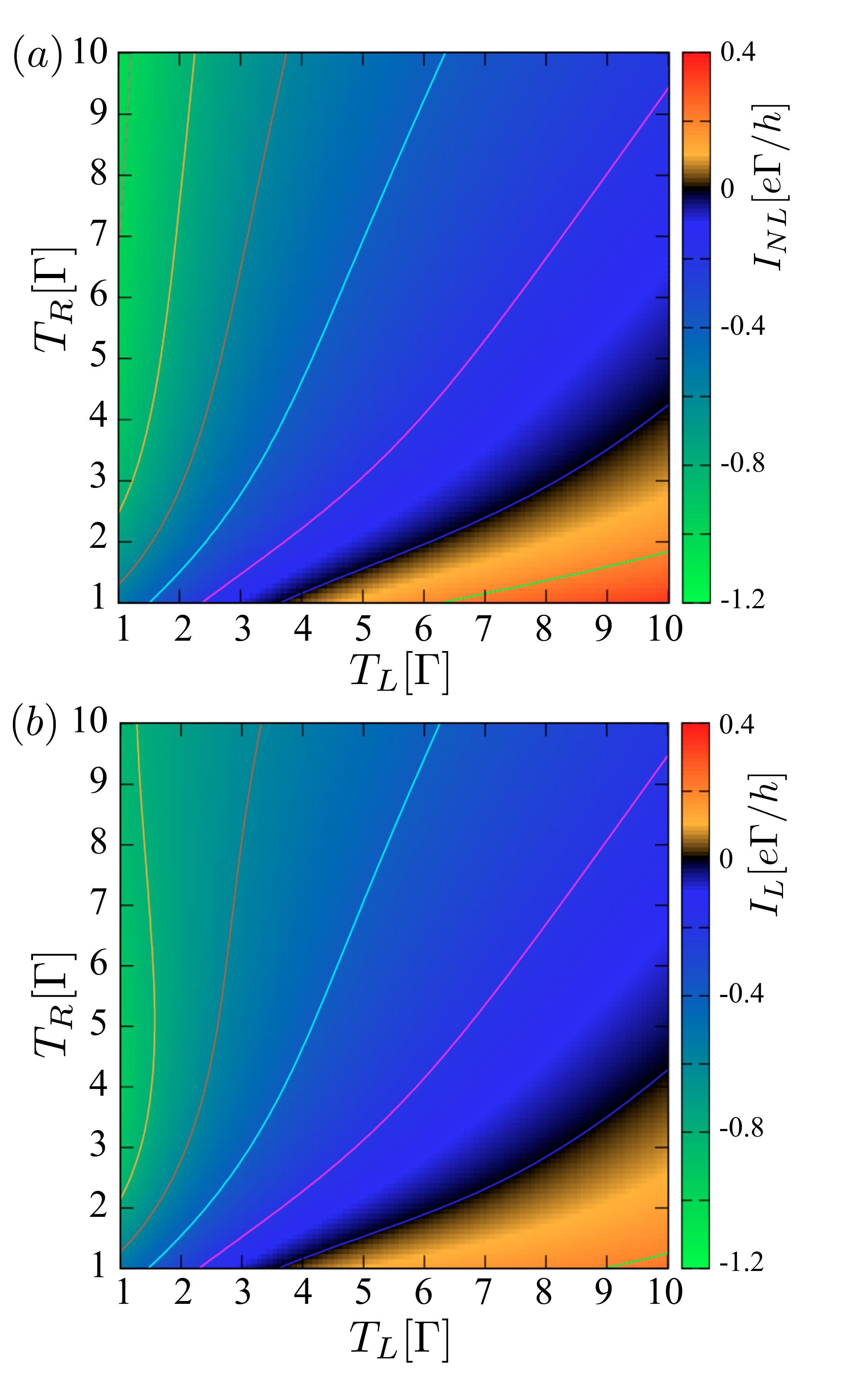}      
      \caption{For the same parameters as in Fig.~\ref{Vbnul}, but for $V_b = 2 \Gamma$, maps of the nonlinear current (a) and linear one (b). The iso-current curves in (a) and (b) are $\sim 0.2 e \Gamma / h$ apart.}
    \label{Vb1}
\end{figure}

\begin{figure}[h!]
            \includegraphics[width=0.5\textwidth]{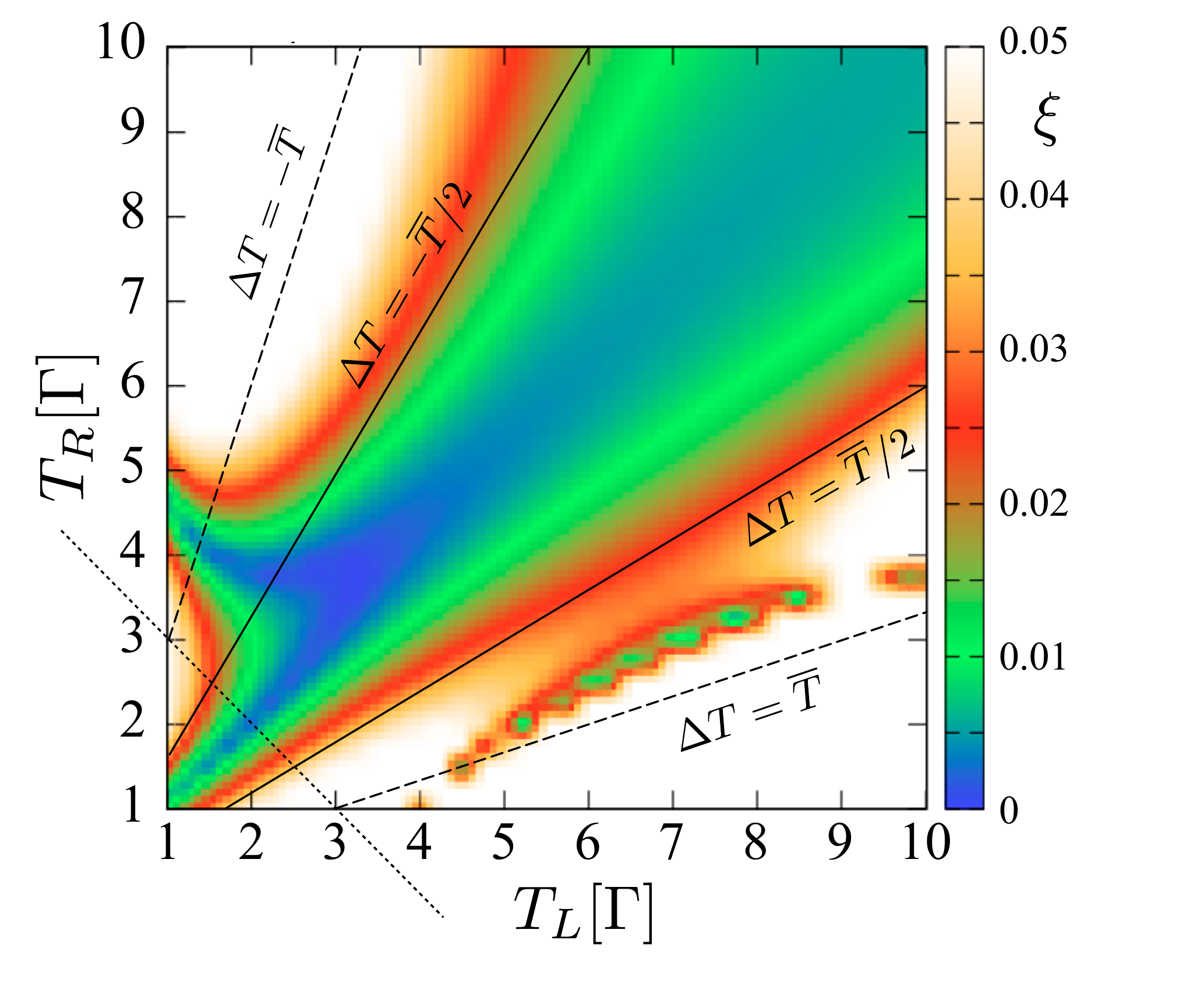}
   \caption{Relative difference between linear and nonlinear currents previously shown in Fig.~\ref{Vb1}. See text for the lines superimposed on the graph.}    
    \label{Vb1bis}
\end{figure}

 Comparisons between linear and nonlinear calculations are presented in Figs.~\ref{Vbnul}-~\ref{Vb1bis}, where 2D plots of the nonlinear current as a function of $T_L$ and $T_R$ are displayed for different voltage biases. 
Maps of the difference between nonlinear and linear currents ($I_{NL}$ and $I_L$) in percent of the nonlinear ones are also shown. 
It may be instructive, looking at the $(T_L, T_R)$ plane, to visualize the diagonal as the $\overline{T}$ axis, while its perpendicular is the $\Delta T$ axis.

In these figures, the main spectral density part contributing to the current is centered close to $\varepsilon_0 + U$. However, the spectral function is the result of an NCA calculation which predicts nonrigid structures.
In Fig.~\ref{Vbnul}, there is no voltage bias, such that the current is purely of thermal origin. According to our convention, it is an odd function of $\Delta T$, 
positive for $\Delta T >0$.
In Fig.~\ref{Vbnul}(b), the cones delimiting the regions $| \Delta T | \le \overline{T} /2$ and $| \Delta T | \le \overline{T}$ have been overlaid on the results. 
They materialize the area where we expect the linear response theory to be reliable
on previous section findings. This map is a quantitative answer to this assumption: inside the cone $| \Delta T | \le \overline{T} /2$ the maximum deviation between linear and nonlinear currents is as low as 3\%, inside the cone $| \Delta T | \le \overline{T} $, it barely exceeds 5\%.

The same kind of current maps and difference between linear and nonlinear currents are shown in the next two figures,  Figs.~\ref{Vb1} and \ref{Vb1bis}, for a finite and moderate voltage bias ($V_b = 2 \Gamma $).
Quantitatively, the same kind of results (not shown) were also obtained  for a higher one ($V_b = 10 \Gamma $).
Due to the voltage bias, the current is no more an odd function of the temperature difference. Given our conventions, and the $\varepsilon_0+U$ value, positive temperature and voltage biases 
are antagonist for the current.
Thus, for the actual bias, the line of vanishing current [black line in Figs.~\ref{Vbnul}(a),  ~\ref{Vb1}(a), and ~\ref{Vb1}(b)] slides from $\Delta T =0$ towards  $\Delta T  >0$. The region where the nonlinear current vanishes is well reproduced by linear calculations as can be seen in Fig.~\ref{Vb1}(b).
In Fig.~\ref{Vb1bis}, in addition to the delimiting temperature cones, a dotted line separating the region where $V_b > \overline{T}$ (left side) from $V_b < \overline{T}$ (right side) has been overlaid on the difference current map. 
The agreement between nonlinear and linear currents, as long as $|V_b| /\overline{T}, |\Delta T |/ \overline{T} \lesssim 1$ is as good as it was in Fig.~\ref{Vbnul}.
It is slightly better on the $\Delta T =T_L -T_R<0$ side of the map for a positive bias, probably for a fortuitous reason: the accordance of the location of spectral density structure, with
places where the Fermi function difference is more safely replaced by its first-order series expansion.

\subsection{Thermopower}

\begin{figure}[h!]
    \includegraphics[width=0.5\textwidth]{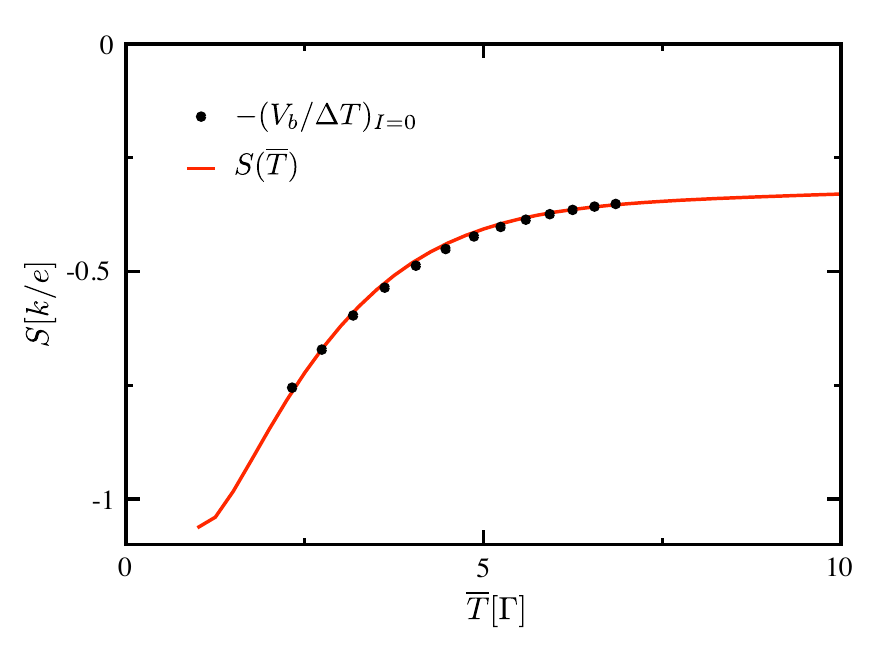}
    \caption{Comparison between the nonlinear thermopower evaluated from the line $I_{NL}=0$ in Fig.~\ref{Vb1}(a), defined as 
    $\Bigl(- V_b / \Delta T\Bigr)_{I_{NL}=0}$, and the linear Seebeck coefficient evaluated at $\overline{T}$.
    }
   \label{curio3}
\end{figure}

In case of non zero voltage bias, 
following the line $I_{NL}=0$ on the map for the nonlinear current in Fig.~\ref{Vb1}(a),  we can evaluate the {\it a priori} nonlinear thermopower 
$S = -\Bigl(\frac{V_b}{\Delta T} \Bigr)_{I_{NL}=0}$. 
This quantity is shown in Fig.~\ref{curio3} as a function of the average temperature $\overline{T}$, and compared to the Seebeck coefficient obtained from equilibrium calculations carried out at $\overline{T}$. For the actual voltage bias, it reveals a remarkable consistency. This is another evidence of the validity of linear approach in this parameter sector. 
Such a successful comparison was already noticed in an
experimental work \cite{Dzurak93}, providing the measure of  the Seebeck coefficient of a quantum point contact. Up to $\Delta T / \overline{T} = 2/3$, the experimental values were successfully compared to a linear evaluation at the operating temperature $\overline{T}$, with a discrepancy less than 10\%.

\section{Kondo Regime}

The NCA is also known to reliably describe the low-temperature Kondo scale $T_{\rm K}$, and to give accurate results for the dot spectral density down to temperatures of the order of a fraction of $T_{\rm K}$~\cite{H97,Aligia2012, NCArecent1}.
We thus investigate the same question concerning the validity of the linear approximation close to the Kondo resonance, still in the case of a doubly degenerate orbital model.
\begin{figure}[h!]
    \includegraphics[width=0.5\textwidth]{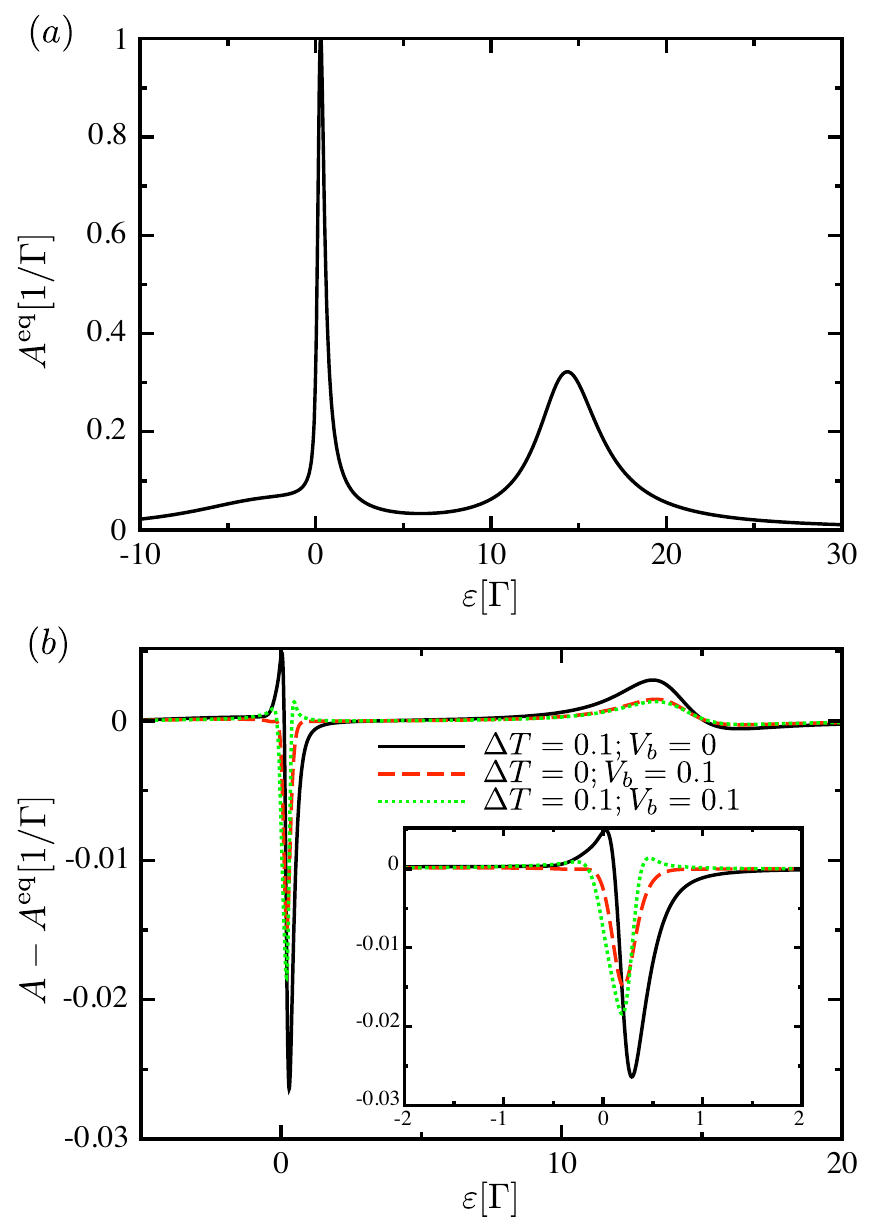}
    \caption{Spin and orbital summed spectral density function close to the Kondo regime for $\varepsilon_0 =-3.2 \Gamma$, $U =16 \Gamma$, $\overline{T}=0.1 \Gamma$ at equilibrium (a), and (b)  difference between non-equilibrium $A$ and equilibrium $A^{\rm eq}$  spectral densities for different thermal and electric biases in $\Gamma$ units. Inset: Close-up view in the same units.}
    \label{Kondo}
\end{figure}
The Kondo physics is restricted to low temperature, more explicitly in quantum dots to low ratio $T/ \Gamma \ll  1$. Moreover, the Kondo structure in the spectral function is known to be readily eroded by a thermal or electrostatic bias \cite{Hershfield91}.
The spectral function at equilibrium is displayed in Fig.~\ref{Kondo}, for $\varepsilon_0 =-3.2 \Gamma$, $U=16 \Gamma$, $\overline{T}=0.1 \Gamma$, together with a plot showing the out-of-equilibrium deviations and some enlarge view of these. For the selected bias values, the narrow peak 
is barely attenuated as expected by $V_b$ and $\Delta T$, whereas in the meantime, to fulfill the spectral function sum rule, the other
structure grows. The peak is also slightly shifted as can be seen in the close-up view of the figure.
Note that the dot density is close to $n \simeq 0.8$, whereas a fully developed Kondo effect is expected  for $n \simeq 1$, then the spectral density would have a maximum height of $4 /(\pi \Gamma)$ at low temperature.
For the actual parameters, the Kondo temperature is about $T_K \simeq 0.5 \Gamma$ \cite{Azema12}.
 The displayed spectral density corresponds to an intermediate regime between Kondo and mixed valence case \cite{Costi10}. The conclusions about  accuracy between linear and nonlinear theory are not sensitive to the dot occupancy, and 
again, as observed previously as long as $|V_b| / \overline{T}, |\Delta T| / \overline{T} \lesssim 1$, the out-of-equilibrium spectral density can be replaced by its equilibrium counterpart without changing substantially  the electric current values.

The validity of the linear approximation  in the Kondo regime was previously noticed and examined in some details in Ref. \cite{Azema12} in the thermogenerator regime. A thorough comparison between linear and nonlinear calculations was done, not only for the electric current, but also for the heat current, with the same conclusions.
In the thermogenerator regime, despite an antagonist voltage, the current flows from the hot to the cold lead. 
For the chosen parameters, the electric current vanishes for a voltage lower than the operating temperature $\overline{T}$. 
It was shown in these conditions, that the output power as well as the device efficiency were qualitatively successfully reproduced, in a wide $(V_b, \Delta T)$ area, and quantitatively 
for $\Delta T \lesssim \overline{T}$.

\section{miscellaneous}

\subsection{Beyond the wide-band-limit}

It is common in transport theory through a central channel to use a smooth density of states for both leads, such that the transport properties will be found to depend little on their specific shapes.
In the previous calculations, we used for each lead a Gaussian function for $\Gamma_{\alpha}(\varepsilon)$ characterized by a full width at half-maximum $2D =  169 \Gamma$, by far the largest energy scale of the system. 
To question the robustness of our findings concerning the domain of validity of the linear response theory, 
for the same parameters than those used in Fig.~\ref{Vbnul},  we have calculated the nonlinear current and evaluated the relative difference between linear and nonlinear ones, in case of a lead density of state built from a narrower Gaussian function, characterized by $2D = 3.3 \Gamma$. Due the shrinkage of $\Gamma(\varepsilon)$, the current is reduced by a factor of roughly three. The area of relevance of the linear theory based on the same criteria of 5\% is not very different from what we had in Fig.~\ref{Vbnul}(b), just slightly narrower.

\subsection{ Affine function and linear theory}

We conclude this study by a calculation showing that linear transport theory can account for a nonmonotonic bias dependence.
It is common in experimental setups to fix the {\sl cold} reservoir temperature.
Thus the increase of  $\Delta T$ leads to a drift of the operating temperature. As a consequence, the linear transport coefficients evaluated at $\overline{T}$ vary and 
the current is not {\it a priori} an affine function of $\Delta T$.
Had the temperature bias been symmetrically applied, $\overline T$ would have been fixed, and the linear current would have been an affine function of the temperature difference.

\begin{figure}[h!]
    \includegraphics[width=0.5\textwidth]{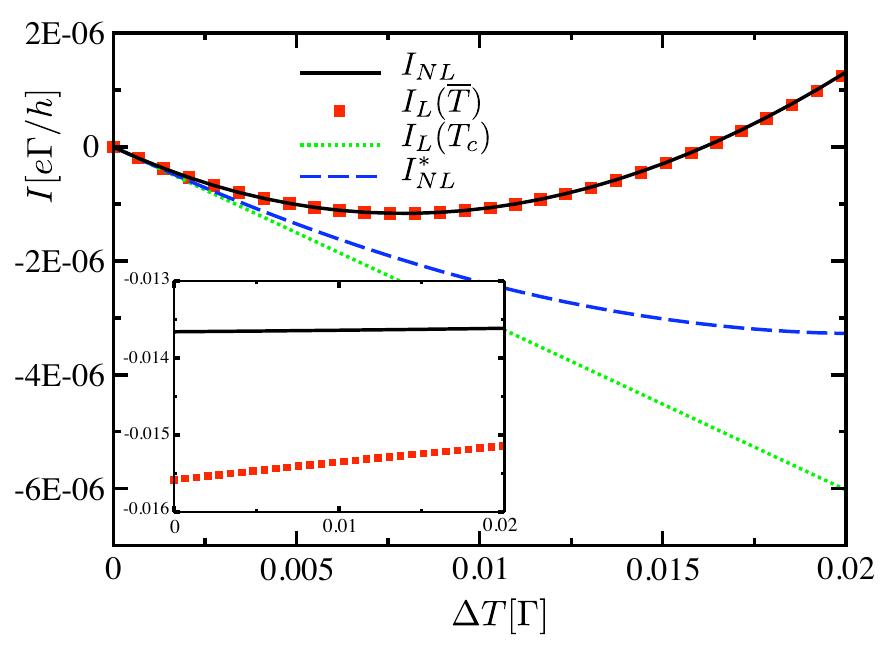}
    \caption{Nonlinear current $I_{NL}$ for $\varepsilon_0 = -37.42 \Gamma, U=47.77 \Gamma, V_b=0$, and $T_C = 4 \ 10^{-2} \Gamma$, as a function of thermal bias. It is compared to three approximations: linear approach at the operating temperature $I_L(\overline{T})$, linear approximation at the fixed low temperature $I_L(T_C)$, and nonlinear calculation with a rigid spectral function evaluated at $T_C$, $I_{NL}^*$.
    Inset: same parameters as previously, except $V_b= 0.4 \Gamma$. }
   \label{curio1}
\end{figure}

The calculation is done in a regime of evanescent Kondo physics: a low-energy structure appears in the gap, created by Kondo fluctuations.
For the following parameters: $\varepsilon_0 = -37.42 \Gamma, U=47.77 \Gamma, V_b=0$, and a fixed cold-side temperature $T_C = 4 \times 10^{-2} \Gamma$, the nonlinear current 
$I_{NL}$ is plotted in Fig.~\ref{curio1} as a function of $\Delta T =T_H - T_C$ ($T_H$ is the temperature of the {\it hot} lead). The origin of the current is purely thermal
(with a very small value compared to previous examples, due to a faint spectral density in the window defined by the temperature bias). $I_{NL}$
is a nonmonotonic function of $\Delta T$. This nonmonotonous behavior is closely related to the spectral function evolution as a function of temperatures as shown soon.
The nonlinear current cannot be distinguished from the calculation made in a linear approach, if carried out at the operating temperature $\overline{T} = (T_C +T_H)/2$; while the linear theory made at the fixed temperature $T_C$ simply predicts an affine relation between $I$ and $\Delta T$, with the slope being equal to $-S(T_c)G(T_c)$. 
Note that the local slope of $I_{L}(\overline{T})$ is related to $S(\overline{T})G(\overline{T})$, which changes with $\Delta T$, even its sign changes.

Finally, to highlight the temperature dependence of the spectral density, and its influence on the current value and even direction, we added on this plot $I_{NL}^{*}$, the current obtained from the integral of the exact Fermi function difference $(f_L-f_R)$, times the  spectral function evaluated at $T_C$. This enables to disentangle the role played by the Fermi function difference from the role played by the transfer function.
Neither $I_L(T_C)$, nor $I_{NL}^{*}$ reproduces the nonlinear current.
On this plot, the operating temperature scans the range $\overline{T} = T_C +\Delta T /2 \in [4 \times 10^{-2}\Gamma ,\ 5 \times 10^{-2}\Gamma]$.

Now turning on a voltage bias, for the same parameters, with $V_b= 0.4 \Gamma$  (much higher than the varying $\overline{T}$,
such that the linear transport theory is not expected to be valid), the current, then essentially of electrostatic origin, is larger by several orders of magnitude than without $V_b$.
Despite an affine relation between current and thermal bias, as can be seen in the inset of Fig.~\ref{curio1}, the linear 
transport theory at $\overline{T}$ is unable to reproduce the out-of-equilibrium result.\cite{rem}

\section{conclusion}

By detailed and wide parameter range comparisons, using a technique which treats repulsion between electrons in the out-of-equilibrium situation, we have shown that the linear theory in voltage and thermal biases, which greatly alleviate transport calculations, in presence of correlations, has an unexpected large range of validity. The only criteria are $|V_b| \lesssim \overline{T}$ and $|\Delta T| \lesssim \overline{T}$, and to be relevant, the transport coefficients have to be evaluated at the average or operating temperature $\overline{T}$.
We have shown that $\Gamma$, the hybridization parameter, has no influence on the linear expansion, this was not obvious from the beginning: the accuracy of the replacement of 
$A(\varepsilon, \overline{T},\Delta T, V_b)$ by  $A(\varepsilon, \overline{T},0,0)$ might have depend on $\Gamma /\overline{T}$.
The study was undertaken for a doubly-degenerate model, however, 
our conclusions do not rely on the orbital number.
 Besides, note that the case of dissymmetric coupling to both leads needs a generalization of the NCA code, and is left for further investigation.

The operating or average temperature $\overline{T}$ is not the dot temperature, which cannot be defined in such an out-of-equilibrium situation.
There is no mathematical evidence for such a dependence on $\overline{T}$:  
in NCA calculations\cite{HKH98}, the dot properties, and especially, the dot occupancy, in case of symmetric coupling $\Gamma_L=\Gamma_R$, are calculated using some average of the lead Fermi functions: $\overline{f} =(f_L+f_R)/2$. It is easy to convince oneself that this last function may be quite different from $f(\overline{T},\overline{\mu})$.  

It is experimentally more convenient to keep one reservoir temperature fixed, usually the cold one (for example, see Ref. \cite{Dzurak93}).
Accordingly, the linear behavior is then often explored from this ${T_C}$.
We feel that the widespread claim of nonlinear behavior should be reexamined with the idea of linearization made at operating temperature,
except when the dot level is also artificially temperature-dependent, as supposed in Refs. \cite{Grifoni12, Leijnse2010}; in that case, a linear behavior is excluded from the beginning.
In Ref. \cite{Reddy07}, the temperature and voltage biases are rather low: one temperature is fixed to 300~K, while the bias  $\Delta T $ grows up to 30~K; furthermore 
the bias voltage is as low as $V_b = 300~\mu$V. We thus expect on the grounds of our study, that linear response theory may be reliable. The observed faint
non-linearity could be attributed to the fact that the Anderson model with two orbitals leaded to non correlated reservoirs  may be 
caricatural and inadequate to describe the experimental system.
Experimentally, $\Delta T$ barely reaches $\overline{T}$, while this is common for the electrostatic bias:
in the work presented in Ref. \cite{Svensson13}, the nonlinear regime is readily attained due to a high voltage, while the temperature difference stays lower than the mean temperature.

Our criteria for linear theory relevance, in this paper only presented for the electric current, are also pertinent for the heat current, due to a very similar influence of Fermi function difference and spectral density. For example, the heat current flowing from the left electrode to the dot reads
$ I_{Q_L}= \frac{2}{h}\int  (\varepsilon -\mu_L) \left[f_{\rm L}(\varepsilon)-f_{\rm R}(\varepsilon)\right]\tau(\varepsilon)\,{\rm d}\varepsilon.$
The study presented in Ref. \cite{Azema12}, restricted to the thermogenerator regime, validates the criteria.


\begin{thebibliography}{99}
\bibitem{rem} Neither is the linear transport theory at $T_C$ nor at $T_H$, able to reproduce this affine behavior.
\end{thebibliography}
\end{document}